\documentclass[10pt,ams]{article}
\usepackage{amsfonts}
\usepackage{graphicx}
\usepackage{amsmath}

\begin{document}

\date{}
\title{\textbf{Study of the Gribov region in Euclidean Yang-Mills theories
in the maximal Abelian gauge}}
\author{ \textbf{M.A.L. Capri$^a$\thanks{marcio@dft.if.uerj.br}}\,,
\textbf{A.J.G{\'o}mez$^a$\thanks{ajgomez@uerj.br}}\,,
\textbf{V.E.R.Lemes$^{a}$\thanks{vitor@dft.if.uerj.br}}\,,
 \\\textbf{R.F. Sobreiro}$^{a}$\thanks{%
sobreiro@uerj.br}\,, \textbf{S.P. Sorella}$^{a}$\thanks{%
sorella@uerj.br}\;,\footnote{Work supported by FAPERJ, Funda{\c
c}{\~a}o de Amparo {\`a} Pesquisa do Estado do Rio de Janeiro,
under the program {\it Cientista do Nosso Estado},
E-26/100.615/2007.}   \\\\
\textit{$^{a}$\small{UERJ
$-$ Universidade do Estado do Rio de Janeiro}}\\
\textit{\small{Instituto de F\'{\i }sica $-$ Departamento de
F\'{\i
 }sica Te\'{o}rica}}\\
\textit{\small{Rua S{\~a}o Francisco Xavier 524, 20550-013
Maracan{\~a}, Rio de Janeiro, Brasil}} \\ } \maketitle

\begin{abstract}
The properties of the Gribov region in $SU(2)$ Euclidean
Yang-Mills theories in the maximal Abelian gauge are investigated.
This region turns out to be bounded in all off-diagonal
directions, while it is unbounded along the diagonal one. The soft
breaking of the BRST\ invariance due to the restriction of the
domain of integration in the path integral to the Gribov region is
scrutinized. Owing to the unboundedness in the diagonal direction,
the invariance with respect to Abelian transformations is
preserved, a property which is at the origin of the local $U(1)$
Ward identity of the maximal Abelian gauge.


\end{abstract}

\newpage


\section{Introduction}

In recent years, the maximal Abelian gauge has been largely
employed in order to investigate nonperturbative aspects of
Yang-Mills theories. The dual superconductivity mechanism for
color confinement \cite{nb,md,'tHooft:1981ht}, the Abelian
dominance hypothesis
\cite{Ezawa:bf,Suzuki:1989gp,Suzuki:1992gz,Hioki:1991ai} and the
infrared behavior of the two point gluon and ghost correlation
functions
\cite{Amemiya:1998jz,Bornyakov:2003ee,Mendes:2006kc,Mendes:2008ux,Schaden:1999ew,Kondo:2001nq,
Lemes:2002ey,Dudal:2002xe,Dudal:2004rx,Capri:2005tj,Dudal:2005bk,Capri:2006cz,Capri:2006vv,Capri:2008ak}
are examples of such nonperturbative aspects. \\\\One important
feature of the maximal Abelian gauge is that it possesses a
lattice formulation \cite{Kronfeld:1987vd,Kronfeld:1987ri}, while
being a renormalizable gauge in the continuum
\cite{Min:1985bx,Fazio:2001rm,Dudal:2004rx}, a property which has
provided a useful comparison among results obtained through
numerical simulations and theoretical investigations.
\\\\As far as the gluon and ghost propagators are concerned, their
study in the maximal Abelian gauge has followed a pattern
analogous to that employed\ in the case of the Landau gauge
\cite{Gribov:1977wm,Zwanziger:1989mf,Zwanziger:1992qr}. Due to the
existence of the Gribov copies \cite{Bruckmann:2000xd}, the
allowed gauge field configurations are restricted to the Gribov region $%
\Omega $, defined as the set of field configurations corresponding
to all relative minima of the minimizing functional
$\mathcal{F}[A]$ \cite{Bruckmann:2000xd}, given by
\begin{equation}
\mathcal{F}[A]=\int {d^{4}x}\text{ }A_{\mu }^{a}A_{\mu }^{a}\ ,  \label{i1}
\end{equation}%
where the index $a=1,2$ runs over the off-diagonal components of
the gauge field. \\\\In particular, the restriction to the region
$\Omega $ in the Feynman path integral has been achieved by
following the framework outlined by Zwanziger in the Landau gauge
\cite{Zwanziger:1989mf,Zwanziger:1992qr}, amounting to add to the
Yang-Mills action a nonlocal term, known as the horizon term.
Albeit nonlocal, the horizon term can be cast in local form
through the introduction of a set of auxiliary fields, leading to
a local action which enjoys the property of being renormalizable
\cite{Capri:2006cz,Capri:2006vv,Capri:2008ak}. This is the
starting point for the analytic investigation of the gluon and
ghost propagators. We underline that the results obtained so far
\cite{Capri:2008ak} display a remarkable agreement with the most
recent lattice data \cite{Mendes:2006kc,Mendes:2008ux}, see Sect.4
for a brief review.
\\\\Though, our current knowledge of the properties of the Gribov
region in the maximal Abelian gauge has not yet reached the same
understanding which has been achieved in the case of the Landau
gauge \cite{Gribov:1977wm,Dell'Antonio:1991xt,vanBaal:1991zw}. A
better knowledge of this region would be of great help in order to
investigate the nonperturbative behavior of nonabelian gauge
theories quantized in the maximal Abelian gauge. \\\\This work
aims at filling part of this gap. We shall establish a few results
on the Gribov region in the maximal Abelian gauge,  providing a
better understanding of several features displayed by this gauge.
This will be the case, for example, of the existence of a local
$U(1)$ Ward identity which has a natural interpretation within the
Abelian dominance hypothesis, according to which the relevant
degrees of freedom at low energies should correspond to those
encoded in the diagonal component of the gauge field. The
off-diagonal components are expected to develop a dynamical mass
which decouple them in the low momentum region, a feature which
has received support from both lattice
\cite{Amemiya:1998jz,Bornyakov:2003ee,Mendes:2006kc,Mendes:2008ux}
and analytic investigations \cite{Dudal:2004rx}. \\\\More
specifically, we shall see that the Gribov region $\Omega $ of the
maximal Abelian gauge turns out to be bounded in all off-diagonal
directions in field space, while it is unbounded in the diagonal
one. This feature makes the Gribov region of the maximal Abelian
gauge different from that of the Landau gauge, which is known to
be bounded in all directions. Moreover, the unboundedness along
the diagonal direction turns out to be at the origin of the $U(1)$
local Ward identity, which holds even in the presence of the
horizon function implementing the restriction to the region
$\Omega $. The convexity of the Gribov region $\Omega $ will be
also established. Furthermore, as in the case of the Landau gauge
\cite{Dudal:2008sp}, $BRST$ invariance turns out to be softly
broken by the presence of the Gribov horizon in the off-diagonal
directions. As we shall see, this breaking originates from the
fact that any infinitesimal gauge transformation of the
off-diagonal field components gives rise to field configurations
lying outside of the Gribov region. \\\\The paper is organized as
follows. Sect.2 is devoted to the study of the Gribov region
$\Omega$ in the maximal Abelian gauge. After establishing that the
Gribov region is bounded in the off-diagonal directions and
unbounded along the diagonal one, we shall face the issue of the
convexity of $\Omega$. Also, Gribov statement's about
infinitesimal copies located near the horizon will be employed to
establish that any infinitesimal gauge transformation of the
off-diagonal components of a gauge configuration belonging to
$\Omega $ will give rise to a field configuration lying outside of
$\Omega $. In Sect.3 we revise the introduction of the horizon
function and we discuss the issues of the the soft breaking of the
$BRST$ invariance and of the $U(1)$  local Ward identity, in the
light of the aforementioned properties of the Gribov region.
Sect.4 contains a brief survey of the main results obtained for
the gluon and ghost propagators. In Sect.5 we present our
conclusion.

\section{Properties of the Gribov region in the maximal Abelian gauge}

\subsection{Gauge fixing conditions}

In this section we discuss the gauge fixing conditions. Let us begin with
the standard notation employed in the case of the maximal Abelian gauge. The
gauge field $\mathcal{A}_{\mu }$ is decomposed as
\begin{equation}
\mathcal{A}_{\mu }=A_{\mu }^{A}T^{A}\equiv A_{\mu }^{a}T^{a}+A_{\mu }T^{3}\;.
\label{n1}
\end{equation}%
where $T^{3}$ stands for the diagonal generator of the $U(1)$ Cartan
subgroup of $SU(2)$, while the index $a=1,2$ labels the remaining
off-diagonal generators $\{T^{a}\}$. \ Similarly to the decomposition of the
gauge field $\mathcal{A}_{\mu }$, for the field strength one has
\begin{equation}
\mathcal{F}_{\mu \nu }=F_{\mu \nu }^{a}T^{a}+F_{\mu \nu }T^{3}\;,  \label{n2}
\end{equation}%
with the off-diagonal and diagonal components given by
\begin{eqnarray}
F_{\mu \nu }^{a} &=&D_{\mu }^{ab}A_{\nu }^{b}-D_{\nu }^{ab}A_{\mu }^{b}\;,
\notag \\
F_{\mu \nu }^{3} &\equiv &F_{\mu \nu }=\partial _{\mu }A_{\nu }-\partial
_{\nu }A_{\mu }+g\varepsilon ^{ab}A_{\mu }^{a}A_{\nu }^{b}\;,  \notag \\
\varepsilon ^{ab} &\equiv &\varepsilon ^{3ab}\;,  \label{not}
\end{eqnarray}%
where we have introduced the covariant derivative $D_{\mu }^{ab}$ with
respect to the diagonal components $A_{\mu }$ of the gauge field, namely
\begin{equation}
D_{\mu }^{ab}\equiv \delta ^{ab}\partial _{\mu }-g\varepsilon ^{ab}A_{\mu
}\;.  \label{covdev}
\end{equation}%
For the Yang-Mills action in Euclidean space one obtains
\begin{equation}
S_{\mathrm{YM}}=\frac{1}{4}\int d^{4}x\,\left( F_{\mu \nu }^{a}F_{\mu \nu
}^{a}+F_{\mu \nu }F_{\mu \nu }\right) \;.  \label{ym}
\end{equation}%
As it is easily checked, the classical action (\ref{ym}) is left invariant
by the gauge transformations
\begin{eqnarray}
\delta A_{\mu }^{a} &=&-D_{\mu }^{ab}{\omega }^{b}-g\varepsilon ^{ab}A_{\mu
}^{b}\omega \;,  \notag \\
\delta A_{\mu } &=&-\partial _{\mu }{\omega }-g\varepsilon ^{ab}A_{\mu
}^{a}\omega ^{b}\;,  \label{gauge}
\end{eqnarray}%
where $\left( {\omega }^{a},{\omega }\right) $ stand for the
off-diagonal and diagonal infinitesimal gauge parameters,
respectively. The maximal Abelian gauge is obtained by demanding
that the off-diagonal components $ A_{\mu }^{a}$ of the gauge
field obey the nonlinear condition
\begin{equation}
D_{\mu }^{ab}A_{\mu }^{b}=0\;,  \label{offgauge}
\end{equation}%
which follows by requiring that the auxiliary functional
\begin{equation}
\mathcal{F}[A]=\int {d^{4}x}A_{\mu }^{a}A_{\mu }^{a}\;,  \label{fmag}
\end{equation}%
is stationary with respect to the gauge transformations (\ref{gauge}).
Moreover, as it is apparent from the presence of the covariant derivative $%
D_{\mu }^{ab}$, equation (\ref{offgauge}) allows for a residual local $U(1)$
invariance corresponding to the diagonal subgroup of $SU(2)$. This
additional invariance has to be fixed by means of a suitable gauge condition
on the diagonal component $A_{\mu }$, which will be chosen to be of the
Landau type, also adopted in lattice simulations, namely
\begin{equation}
\partial _{\mu }A_{\mu }=0\;.  \label{dgauge}
\end{equation}

\subsection{The Gribov region of the maximal Abelian gauge}

In order to introduce the Gribov region $\Omega $ in the maximal
Abelian gauge, let us first remind a few properties of the
Faddeev-Popov operator, $ \mathcal{M}^{ab}$, which is obtained by
taking the second variation of the auxiliary functional
$\mathcal{F}[A]$:
\begin{equation}
\delta^2 \mathcal{F}[A] = 2 \int {d^{4}x}\; \omega^a
\mathcal{M}^{ab} \omega^b  \ , \label{sv}
\end{equation}
where
\begin{equation}
\mathcal{M}^{ab}=-D_{\mu }^{ac}D_{\mu }^{cb}-g^{2}\varepsilon
^{ac}\varepsilon ^{bd}A_{\mu }^{c}A_{\mu }^{d}\;.  \label{offop}
\end{equation}%
The operator $\mathcal{M}^{ab}$ enjoys the property of being
Hermitian and, as pointed out in \cite{Bruckmann:2000xd}, is the
difference of two positive semi-definite operators
$\emph{O}_{1}^{ab}$ and $\emph{O}_{2}^{ab}$, namely
\begin{eqnarray}
\mathcal{M}^{ab} &=&\emph{O}_{1}^{ab}-\emph{O}_{2}^{ab}\;,  \notag \\
\emph{O}_{1}^{ab} &=&-D_{\mu }^{ac}D_{\mu }^{cb}\;,  \notag \\
\emph{O}_{2}^{ab} &=&g^{2}\varepsilon ^{ac}\varepsilon ^{bd}A_{\mu
}^{c}A_{\mu }^{d}\newline =g^{2}\widetilde{A}_{\mu
}^{a}\widetilde{A}_{\mu }^{b}\;,  \label{gr1}
\end{eqnarray}%
where we have introduced the notation $\widetilde{A}_{\mu }^{a}=\varepsilon
^{ac}A_{\mu }^{c}$. The positivity of both operators $\emph{O}_{1}^{ab}$ and
$\emph{O}_{2}^{ab}$ is easily established. In fact
\begin{equation}
\left\langle \psi \left\vert \emph{O}_{1}\right\vert \psi
\right\rangle =-\int d^{4}x\;\left( \psi ^{a}\right) ^{\dagger
}D_{\mu }^{ac}D_{\mu }^{cb}\psi ^{b}=\int d^{4}x\;\left( D_{\mu
}^{ac}\psi ^{c}\right) ^{\dagger }\left( D_{\mu }^{ab}\psi
^{b}\right) =\left\Vert D_{\mu }^{ab}\psi ^{b}\right\Vert ^{2}\geq
0\;.  \label{gr2}
\end{equation}%
\newline
Analogously
\begin{equation}
\left\langle \psi \left\vert \emph{O}_{2}\right\vert \psi \right\rangle
=\int d^{4}x\;\left( \psi ^{a}\right) ^{\dagger }g^{2}\widetilde{A}_{\mu
}^{a}\widetilde{A}_{\mu }^{b}\psi ^{b}=\left\Vert g\widetilde{A}_{\mu
}^{a}\psi ^{a}\right\Vert ^{2}\geq 0\;.  \label{gr3}
\end{equation}%
\newline
It is worth noticing that the operator $\emph{O}_{1}$ depends only on the
diagonal component $A_{\mu }$, $\emph{O}_{1}=\emph{O}_{1}(A)$, while $\emph{O%
}_{2}$ contains only the off-diagonal fields $A_{\mu }^{a}$, $\emph{O}_{2}=%
\emph{O}_{2}(A_{\mu }^{a})$. \\\\As in the case of the Landau
gauge \cite{Zwanziger:1989mf,Zwanziger:1992qr}, the Gribov region
$\Omega $ of the maximal Abelian gauge is defined as the set of
all relative minima of the auxiliary functional $\mathcal{F}[A]$,
being given by the set of fields fulfilling the gauge conditions
(\ref{offgauge}), (\ref{dgauge}), and for which the Faddeev-Popov
operator $\mathcal{M}^{ab}$ is positive definite, namely
\begin{equation}
\Omega =\left\{ A_{\mu },\;A_{\mu }^{a},\;\partial _{\mu }A_{\mu
}=0,\;D_{\mu }^{ab}A_{\mu }^{b}=0,\;\mathcal{M}^{ab}=-D_{\mu }^{ac}D_{\mu
}^{cb}-g^{2}\varepsilon ^{ac}\varepsilon ^{bd}A_{\mu }^{c}A_{\mu
}^{d}>0\right\} \;.  \label{gr4}
\end{equation}%
In the following, a few properties of the  region $\Omega $ will
be established.

\subsubsection{Properties of the region $\Omega $ along the off-diagonal directions}

\begin{itemize}
\item \textbf{Statement}: \textit{The Gribov region }$\Omega $\textit{\ is
bounded in all off-diagonal directions}
\end{itemize}

\noindent In order to prove this statement we observe that if
$\left( B_{\mu }^{a},B_{\mu }\right) $ is a field configuration
fulfilling the maximal Abelian gauge conditions, $D_{\mu
}^{ab}B_{\mu }^{b}=0,\,\partial _{\mu }B_{\mu }=0$, then the
re-scaled configuration $\left( \lambda B_{\mu }^{a},B_{\mu
}\right) $, with $\lambda $ a positive constant factor, obeys the
same gauge condition. In fact
\begin{equation}
D_{\mu }^{ab}\left( \lambda B_{\mu }^{b}\right) =\lambda D_{\mu }^{ab}B_{\mu
}^{b}=0\;.  \label{gr5}
\end{equation}%
Let now $\left( A_{\mu }^{a},A_{\mu }\right) $ be a field configuration
belonging to $\Omega $, \textit{i.e.}
\begin{equation}
\left\langle \psi \left\vert \mathcal{M}(A_{\mu }^{a},A_{\mu })\right\vert
\psi \right\rangle =\left\langle \psi \left\vert \emph{O}_{1}(A)\right\vert
\psi \right\rangle -\left\langle \psi \left\vert \emph{O}_{2}(A_{\mu
}^{a})\right\vert \psi \right\rangle >0\;.  \label{gr6}
\end{equation}%
Let us consider the re-scaled configuration $\left( \lambda A_{\mu
}^{a},A_{\mu }\right) $ and let us evaluate $\left\langle \psi
\left\vert \mathcal{M}(\lambda A_{\mu }^{a},A_{\mu })\right\vert
\psi \right\rangle $, namely
\begin{equation}
\left\langle \psi \left\vert \mathcal{M}(\lambda A_{\mu }^{a},A_{\mu
})\right\vert \psi \right\rangle =\left\langle \psi \left\vert \emph{O}%
_{1}(A)\right\vert \psi \right\rangle -\lambda ^{2}\left\langle \psi
\left\vert \emph{O}_{2}(A_{\mu }^{a})\right\vert \psi \right\rangle \;.
\label{gr7}
\end{equation}%
Since both $\left\langle \psi \left\vert \emph{O}_{1}(A)\right\vert \psi
\right\rangle $ and $\left\langle \psi \left\vert \emph{O}_{2}(A_{\mu
}^{a})\right\vert \psi \right\rangle $ are positive definite, it follows
that for $\lambda $ large enough the right hand side of eq.(\ref{gr7}) will
become negative, meaning that one has left the Gribov region $\Omega $. This
shows that moving along the off-diagonal directions parametrized by the
re-scaled configuration $\left( \lambda A_{\mu }^{a},A_{\mu }\right) $, with $%
\left( A_{\mu }^{a},A_{\mu }\right) $ belonging to the Gribov region $\Omega
$, one always encounters a boundary $\partial \Omega $, \textit{i.e.} the
horizon, where the first vanishing eigenvalue of the Faddeev-Popov operator
appears. Beyond $\partial \Omega $, the operator $\mathcal{M}^{ab}$ ceases
to be positive definite.

\subsubsection{Unboundedness of $\Omega $ in the diagonal direction}

\begin{itemize}
\item \textbf{Statement:} \textit{The region }$\Omega $\textit{\ is
unbounded in the diagonal direction}
\end{itemize}

\noindent To prove this statement it is sufficient to observe that
the purely diagonal field configuration $\left(
\overrightarrow{0},A_{\mu }\right) $ with $A_{\mu }$ transverse,
$\partial _{\mu }A_{\mu }=0$, fulfills the maximal Abelian gauge
condition. Moreover, for this kind of configuration, the
Faddeev-Popov operator $\mathcal{M}^{ab}$ reduces to the covariant
Laplacian
\begin{equation}
\mathcal{M}^{ab}(\overrightarrow{0},A_{\mu })=-D_{\mu }^{ac}(A)D_{\mu
}^{cb}(A)\;,  \label{g7b}
\end{equation}%
which is always positive for an arbitrary choice of the transverse
diagonal configuration $A_{\mu }$. We see thus that one can freely
move along the diagonal direction in field space. The Faddeev-Popov operator $\mathcal{M}%
^{ab}$ will never become negative, meaning that the region $\Omega $ is
unbounded in the diagonal direction.

\subsubsection{Convexity of the region $\Omega $}

Let us face now the issue of the convexity of the region $\Omega
$. Due to the nonlinearity of the gauge conditions, this property
will be established for configurations lying on the same diagonal
hyperplane in field space. Let us consider in fact two field configuration $%
\left( B_{\mu }^{a},A_{\mu }\right) $, $\left( C_{\mu }^{a},A_{\mu }\right) $
fulfilling the gauge conditions, \textit{i.e.}

\begin{equation}
D_{\mu }^{ab}(A)B_{\mu }^{b}=0\ ,\ \ \ \ D_{\mu }^{ab}(A)C_{\mu }^{b}=0\ ,\
\ \ \ \partial _{\mu }A_{\mu }=0\ ,  \label{gr8}
\end{equation}%
and belonging to the Gribov region $\Omega $
\begin{equation}
\mathcal{M}(B_{\mu }^{a},A_{\mu })>0\ ,\ ~\mathcal{M}(C_{\mu }^{a},A_{\mu
})>0\ .  \label{gr9}
\end{equation}%
Thus, it turns out that the field configuration $\left( E_{\mu
}^{a},A_{\mu }\right) $:
\begin{equation}
E_{\mu }^{a}=\alpha B_{\mu }^{a}+(1-\alpha )C_{\mu }^{a}\ ,\ \ \ \ \ \ \ \
0\leq \alpha \leq 1\ ,  \label{gr10}
\end{equation}%
belongs to $\Omega $, namely
\begin{equation}
\mathcal{M}^{ab}(E_{\mu }^{c},A_{\mu })\ >\ 0\ .  \label{grc}
\end{equation}

\textit{Proof}%
\begin{eqnarray}
\mathcal{M}^{ab}(E_{\mu }^{c},A_{\mu }) &=&-D_{\mu }^{ac}(A)D_{\mu
}^{cb}(A)-g^{2}\widetilde{E}_{\mu }^{a}\widetilde{E}_{\mu }^{b}\;  \notag \\
&=&-D_{\mu }^{ac}(A)D_{\mu }^{cb}(A)-\alpha ^{2}g^{2}\widetilde{B}_{\mu }^{a}%
\widetilde{B}_{\mu }^{b}-(1-\alpha )^{2}g^{2}\widetilde{C}_{\mu }^{a}%
\widetilde{C}_{\mu }^{b}-\alpha (1-\alpha )g^{2}\left( \widetilde{B}_{\mu
}^{a}\widetilde{C}_{\mu }^{b}+\widetilde{C}_{\mu }^{a}\widetilde{B}_{\mu
}^{b}\right)  \notag \\
&=&\alpha ^{2}\left( -D_{\mu }^{ac}(A)D_{\mu }^{cb}(A)-g^{2}\widetilde{B}%
_{\mu }^{a}\widetilde{B}_{\mu }^{b}\right) +(1-\alpha )^{2}\left( -D_{\mu
}^{ac}(A)D_{\mu }^{cb}(A)-g^{2}\widetilde{C}_{\mu }^{a}\widetilde{C}_{\mu
}^{b}\right)  \notag \\
&&+\alpha (1-\alpha )\left( -2D_{\mu }^{ac}(A)D_{\mu }^{cb}(A)-g^{2}\left(
\widetilde{B}_{\mu }^{a}\widetilde{C}_{\mu }^{b}+\widetilde{C}_{\mu }^{a}%
\widetilde{B}_{\mu }^{b}\right) \right) \ .  \label{gr11}
\end{eqnarray}%
From
\begin{equation}
\widetilde{B}_{\mu }^{a}\widetilde{C}_{\mu }^{b}+\widetilde{C}_{\mu }^{a}%
\widetilde{B}_{\mu }^{b}=\widetilde{B}_{\mu }^{a}\widetilde{B}_{\mu }^{b}+%
\widetilde{C}_{\mu }^{a}\widetilde{C}_{\mu }^{b}-\left( \widetilde{C}_{\mu
}^{a}-\widetilde{B}_{\mu }^{a}\right) \left( \widetilde{C}_{\mu }^{b}-%
\widetilde{B}_{\mu }^{b}\right) \ ,  \label{gr12}
\end{equation}%
one has
\begin{eqnarray}
\mathcal{M}^{ab}(E_{\mu }^{c},A_{\mu })\text{ } &=&\alpha ^{2}\mathcal{M}%
^{ab}(B,A)+(1-\alpha )^{2}\mathcal{M}^{ab}(C,A)+\alpha (1-\alpha )\left(
\mathcal{M}^{ab}(B,A)+\mathcal{M}^{ab}(C,A)\right)  \notag \\
&&+\alpha (1-\alpha )g^{2}\left( \widetilde{C}_{\mu }^{a}-\widetilde{B}_{\mu
}^{a}\right) \left( \widetilde{C}_{\mu }^{b}-\widetilde{B}_{\mu }^{b}\right)
\ .  \label{gr13}
\end{eqnarray}%
Since the operator $\emph{O}_{2}^{ab}(C-B)=g^{2}\left( \widetilde{C}_{\mu
}^{a}-\widetilde{B}_{\mu }^{a}\right) \left( \widetilde{C}_{\mu }^{b}-%
\widetilde{B}_{\mu }^{b}\right) \ $is positive definite%
\begin{equation}
\left\langle \psi \left\vert \emph{O}_{2}(C-B)\right\vert \psi \right\rangle
=\left\Vert g\left( \widetilde{C}_{\mu }^{a}-\widetilde{B}_{\mu }^{a}\right)
\psi ^{a}\right\Vert ^{2}\geq 0\;,  \label{gr14}
\end{equation}%
it follows that
\begin{equation}
\mathcal{M}^{ab}(E_{\mu }^{c},A_{\mu })>0\ ,  \label{gr15}
\end{equation}%
showing that the field configuration $\left( E_{\mu }^{a},A_{\mu }\right) $
belongs to $\Omega $, thus establishing the convexity of $\Omega .$

\subsubsection{A statement about field configurations belonging to the
Gribov region and infinitesimal gauge transformations}

In this section we shall discuss how infinitesimal gauge
transformations affect the Gribov region. We shall establish that
any infinitesimal gauge transformation of a field configuration
lying within the region $\Omega $ will give rise to a
configuration which is located outside of $\Omega $. This property
will be at the origin of the soft breaking of the $BRST\
$invariance of the local action implementing the restriction to
the region $\Omega $. In order to prove this statement we shall
distinguish two cases.

\begin{itemize}
\item {\it First case: the field is not located close to the
boundary $\partial \Omega$}

Let us consider a field configuration $\left( A_{\mu }^{a},A_{\mu
}\right) $ belonging to $\Omega $
\begin{equation}
\partial _{\mu }A_{\mu }=0\;,\;\ \ D_{\mu }^{ab}A_{\mu }^{b}=0\;,\;\ \ \
\mathcal{M}^{ab}(A,A^{c})>0\ ,  \label{gr16}
\end{equation}%
and not located close to the boundary $\partial \Omega $. Let us
consider an infinitesimal gauge transformation of the
configuration $\left( A_{\mu }^{a},A_{\mu }\right) $, namely
\begin{eqnarray}
\widetilde{A}_{\mu }^{a} &=&A_{\mu }^{a}-D_{\mu }^{ab}\omega
^{b}-g\varepsilon ^{ab}A_{\mu }^{b}\omega \ ,  \notag \\
\widetilde{A}_{\mu } &=&A_{\mu }-\partial _{\mu }\omega -g\varepsilon
^{ab}A_{\mu }^{a}\omega ^{b}\ ,  \label{gr17}
\end{eqnarray}%
where $\omega ^{a}$ are the off-diagonal components of the
infinitesimal gauge parameter, while $\omega =\omega ^{3}$ is the
diagonal component. Suppose now that $\left( \widetilde{A}_{\mu
}^{a},\widetilde{A}_{\mu }\right) $ belongs to the region $\Omega
$. Thus, we should have
\begin{equation}
\partial _{\mu }\widetilde{A}_{\mu }=0\ ,  \label{gr18}
\end{equation}%
and
\begin{equation}
D_{\mu }^{ab}(\widetilde{A})\widetilde{A}_{\mu }^{b}=0\ .  \label{gr19}
\end{equation}%
From condition (\ref{gr18}) we would get%
\begin{equation}
\partial ^{2}\omega =-g\varepsilon ^{ab}\partial _{\mu }\left( A_{\mu
}^{a}\omega ^{b}\right) \ ,  \label{gr20}
\end{equation}%
while from (\ref{gr19}) it would follow%
\begin{eqnarray}
0 &=&\partial \widetilde{A}_{\mu }^{a}-g\varepsilon ^{ab}\widetilde{A}_{\mu }%
\widetilde{A}_{\mu }^{b}  \notag \\
&=&\partial _{\mu }A_{\mu }^{a}-g\varepsilon ^{ab}A_{\mu }A_{\mu
}^{b}+g\varepsilon ^{ab}A_{\mu }D_{\mu }^{bc}\omega ^{c}+g^{2}\varepsilon
^{ab}A_{\mu }\varepsilon ^{bc}A_{\mu }^{c}\omega  \notag \\
&&+g\varepsilon ^{ab}A_{\mu }^{b}\partial _{\mu }\omega +g^{2}\varepsilon
^{ab}\varepsilon ^{mn}A_{\mu }^{m}\omega ^{n}A_{\mu }^{b}\ -\ \partial _{\mu
}D_{\mu }^{ab}\omega ^{b}-g\varepsilon ^{ab}\partial _{\mu }\left( A_{\mu
}^{b}\omega \right) \   \notag \\
&=&-D_{\mu }^{ac}D_{\mu }^{cb}\omega ^{b}-g^{2}\varepsilon ^{ac}\varepsilon
^{bd}A_{\mu }^{c}A_{\mu }^{d}\omega ^{b}\mathrm{\ ,}  \label{gr21}
\end{eqnarray}%
where terms of higher orders in the infinitesimal parameters $\left( \omega
^{a},\omega \right) $ have been neglected. Therefore, condition (\ref{gr19})
would imply that the Faddeev-Popov operator $\mathcal{M}^{ab}$ should
possess a zero mode, \textit{i.e.}
\begin{equation}
\mathcal{M}^{ab}\omega ^{b}=0\ ,  \label{gr22}
\end{equation}%
which contradicts the fact that the configuration $\left( A_{\mu
}^{a},A_{\mu }\right) $ belongs to the Gribov region $\Omega $. As a
consequence, it follows that the gauge transformed configuration $\left(
\widetilde{A}_{\mu }^{a},\widetilde{A}_{\mu }\right) $, eq.(\ref{gr17}), is
located outside of $\Omega $.

\item {\it Second case: the field is located close to the boundary
$\partial \Omega $}

Let us consider now the case in which the field configuration
$\left( A_{\mu
}^{a},A_{\mu }\right) $ lies very close to the boundary of the region $%
\Omega $. Following \cite{Capri:2005tj}, we can parametrize
$\left( A_{\mu }^{a},A_{\mu }\right) $ as
\begin{eqnarray}
A_{\mu }^{a} &=&C_{\mu }^{a}+a_{\mu }^{a}\ ,  \notag \\
A_{\mu } &=&C_{\mu }+a_{\mu }\ ,  \label{gr23}
\end{eqnarray}%
where $\left( C_{\mu }^{a},C_{\mu }\right) $ lies on the boundary $\partial
\Omega $, namely
\begin{equation}
\partial _{\mu }C_{\mu }=0\ ,\ \ \ \ \ \ D_{\mu }^{ab}(C)C_{\mu }^{b}=0\ ,
\label{gr24}
\end{equation}%
and
\begin{equation}
\mathcal{M}^{ab}(C_{\mu },C_{\mu }^{c})\varphi ^{b}=0\ ,  \label{gr25}
\end{equation}%
where $\varphi ^{b}$ is the zero mode of the Faddeev-Popov operator $%
\mathcal{M}^{ab}(C_{\mu },C_{\mu }^{c})$. The components $\left( a_{\mu
}^{a},a_{\mu }\right) $ in eq.(\ref{gr23}) stand for small perturbations.
Let us also introduce, for later convenience, the quantity $\varphi $
defined as
\begin{equation}
\varphi =-g\varepsilon ^{ab}\frac{1}{\partial ^{2}}\partial _{\mu }\left(
C_{\mu }^{a}\varphi ^{b}\right) \ ,  \label{gr26}
\end{equation}%
so that
\begin{equation}
\partial ^{2}\varphi =-g\varepsilon ^{ab}\partial _{\mu }\left( C_{\mu
}^{a}\varphi ^{b}\right) \ .  \label{gr27}
\end{equation}%
From the gauge conditions%
\begin{equation}
\partial _{\mu }A_{\mu }=0,\;\ \ D_{\mu }^{ab}(A)A_{\mu }^{b}=0\ ,
\label{gr28}
\end{equation}%
it follows that
\begin{eqnarray}
\partial _{\mu }a_{\mu } &=&0\ ,  \notag \\
D_{\mu }^{ab}(C)a_{\mu }^{b}-g\varepsilon ^{ab}a_{\mu }C_{\mu }^{b} &=&0\ ,
\label{gr30}
\end{eqnarray}%
where we have neglected higher order terms in the small components $\left(
a_{\mu }^{a},a_{\mu }\right) $. Performing now an infinitesimal gauge
transformation of the configuration $\left( A_{\mu }^{a},A_{\mu }\right) $,
one gets
\begin{eqnarray}
\widetilde{A}_{\mu }^{a} &=&C_{\mu }^{a}+a_{\mu }^{a}-D_{\mu }^{ab}(C)\omega
^{b}-g\varepsilon ^{ab}C_{\mu }^{b}\omega \ \ ,  \notag \\
\widetilde{A}_{\mu } &=&C_{\mu }+a_{\mu }-\partial _{\mu }\omega
-g\varepsilon ^{ab}C_{\mu }^{a}\omega ^{b}\ .  \label{gr31}
\end{eqnarray}%
We see thus that, unlike the previous case, the new configuration $\left(
\widetilde{A}_{\mu }^{a},\widetilde{A}_{\mu }\right) $ can fulfill the gauge
conditions, provided one identifies the infinitesimal parameters $\left(
\omega ^{a},\omega \right) $ with the components of the zero mode, eqs.(\ref%
{gr25}), (\ref{gr26}) , \textit{i.e. }$\left( \omega ^{a},\omega \right)
=\left( \varphi ^{a},\varphi \right) $. Therefore, the configuration%
\begin{eqnarray}
\widetilde{A}_{\mu }^{a} &=&C_{\mu }^{a}+a_{\mu }^{a}-D_{\mu
}^{ab}(C)\varphi ^{b}-g\varepsilon ^{ab}C_{\mu }^{b}\varphi \ \ ,  \notag \\
\widetilde{A}_{\mu } &=&C_{\mu }+a_{\mu }-\partial _{\mu }\varphi
-g\varepsilon ^{ab}C_{\mu }^{a}\varphi ^{b}\ ,  \label{gr32}
\end{eqnarray}%
obeys the gauge conditions%
\begin{equation}
\partial _{\mu }\widetilde{A}_{\mu }=0,\;\ \ D_{\mu }^{ab}(\widetilde{A})%
\widetilde{A}_{\mu }^{b}=0\ .  \label{gr33}
\end{equation}%
Nevertheless, due to Gribov's statement\footnote{Let us remind
here Gribov's statement, proven in \cite{Gribov:1977wm} in the
case of the Landau gauge, and extended to the maximal Abelian
gauge in \cite{Capri:2005tj} (see Appendix A). Statement: for any
field configuration $\left( A_{\mu }^{a},A_{\mu }\right) $
belonging to the Gribov region $\Omega $ and located
close to the boundary $\partial \Omega$, there exists an equivalent field configuration $%
\left( \widetilde{A}_{\mu }^{a},\widetilde{A}_{\mu }\right) $, given by
\begin{eqnarray}
\widetilde{A}_{\mu }^{a} &=&C_{\mu }^{a}+a_{\mu }^{a}-D_{\mu
}^{ab}(C)\varphi ^{b}-g\varepsilon ^{ab}C_{\mu }^{b}\varphi \ \ ,  \notag \\
\widetilde{A}_{\mu } &=&C_{\mu }+a_{\mu }-\partial _{\mu }\varphi
-g\varepsilon ^{ab}C_{\mu }^{a}\varphi ^{b}\ ,  \label{grf}
\end{eqnarray}%
which is, however, located on the other side of the boundary,
outside of the
region $\Omega $.}, the field $\left( \widetilde{A}_{\mu }^{a},%
\widetilde{A}_{\mu }\right) $ lies precisely outside of the region
$\Omega $.
\end{itemize}

\noindent This ends the proof that any infinitesimal gauge
transformation of a field configuration belonging to $\Omega $
gives rise to a configuration which is located outside of $\Omega
$.

\section{Soft breaking of the $BRST$ invariance due to the restriction to
the Gribov region}

This section is devoted to discuss the issue of the $BRST$\
symmetry when implementing the restriction to the Gribov region.
We shall see that, in a way completely analogous to the case of
the Landau gauge \cite{Dudal:2008sp}, the restriction to the
region $\Omega $ entails a soft breaking of the $BRST$\ symmetry
whose origin can be traced back to the fact that any infinitesimal
gauge transformation of a field configuration belonging to $\Omega
$ gives rise to a configuration which lies outside of $\Omega $.

\subsection{The Faddeev-Popov action and its $BRST$ invariance}

Let us start with the Faddeev-Popov action corresponding to the gauge
conditions (\ref{offgauge}), (\ref{dgauge}), namely
\begin{equation}
S_{\text{\textrm{FP}}}=S_{\mathrm{YM}}+S_{\mathrm{MAG}}\;,  \label{zero}
\end{equation}%
where $S_{\mathrm{YM}}$ is the Yang-Mills action, eq.(\ref{ym}), and $S_{%
\mathrm{MAG}}$ stands for the gauge fixing term of the maximal Abelian
gauge, given by
\begin{equation}
S_{\mathrm{MAG}}=\int {d^{4}\!x\,}\left( \,ib^{a}D_{\mu }^{ab}A_{\mu }^{b}-%
\bar{c}^{a}\mathcal{M}^{ab}c^{b}+g\varepsilon ^{ab}\bar{c}^{a}(D_{\mu
}^{bc}A_{\mu }^{c})c+ib\,\partial _{\mu }A_{\mu }+\bar{c}\,\partial _{\mu
}(\partial _{\mu }c+g\varepsilon ^{ab}A_{\mu }^{a}c^{b})\right) \,\;,
\label{MAG_action}
\end{equation}%
where $(b^{a},b)$ are the off-diagonal and diagonal Lagrange multipliers
enforcing the gauge conditions $D_{\mu }^{ab}A_{\mu }^{b}=0$, $\partial
_{\mu }A_{\mu }=0$. The fields $(c^{a},\bar{c}^{a},c,\bar{c})$ are the
off-diagonal and diagonal Faddeev-Popov ghosts, respectively, and $\mathcal{M%
}^{ab}$ denotes the Faddeev-Popov operator of eq.(\ref{offop}). The action $%
\left( \ref{zero}\right) $ is left invariant by the nilpotent $BRST$
transformation
\begin{equation}
\begin{tabular}{cclccl}
$sA_{\mu }^{a}$ & $\!\!\!=\!\!\!$ & $-(D_{\mu }^{ab}c^{b}+g\varepsilon
^{ab}A_{\mu }^{b}c)\,,\qquad $ & $sA_{\mu }$ & $\!\!\!=\!\!\!$ & $-(\partial
_{\mu }c+g\varepsilon ^{ab}A_{\mu }^{a}c^{b})\,,$\vspace{5pt} \\
$sc^{a}$ & $\!\!\!=\!\!\!$ & $g\varepsilon ^{ab}c^{b}c\,,$ & $sc$ & $%
\!\!\!=\!\!\!$ & $\frac{g}{2}\varepsilon ^{ab}c^{a}c^{b}\,,$\vspace{5pt} \\
$s\bar{c}^{a}$ & $\!\!\!=\!\!\!$ & $ib^{a}\,,$ & $s\bar{c}$ & $\!\!\!=\!\!\!$
& $ib\,,$\vspace{5pt} \\
$sb^{a}$ & $\!\!\!=\!\!\!$ & $0\,,$ & $sb$ & $\!\!\!=\!\!\!$ & $0\,,$%
\end{tabular}
\label{brst_fields}
\end{equation}%
\begin{equation}
s^{2}=0\;.  \label{brstn}
\end{equation}%
Notice that the gauge fixing term $\left( \ref{MAG_action}\right) $ can be
written as an exact BRST variation
\begin{equation}
S_{\mathrm{MAG}}=s\int {d^{4}\!x\,}\left( \bar{c}^{a}D_{\mu }^{ab}A_{\mu
}^{b}+\bar{c}\,\partial _{\mu }A_{\mu }\right) \;.  \label{brstex}
\end{equation}

\subsection{Introduction of the horizon function, localization, and
soft breaking of the $BRST$ invariance}

As already mentioned, the maximal Abelian gauge is affected by the
existence of Gribov copies \cite{Bruckmann:2000xd}, which have to
be taken into account in order to properly quantize the theory. To
deal with this problem, it is necessary to restrict the domain of
integration in the Feynman path integral to the Gribov region
$\Omega $. As in the case of the Landau gauge
\cite{Zwanziger:1989mf,Zwanziger:1992qr}, this restriction is
achieved through the introduction of the horizon function $
S_{\mathrm{Hor}}$ which, in the case of the maximal Abelian gauge,
is given by the following nonlocal expression
\cite{Capri:2006cz,Capri:2008ak}
\begin{equation}
S_{\mathrm{Hor}}=\gamma ^{4}g^{2}\int {d^{4}\!x\,}\varepsilon ^{ab}A_{\mu
}\left( \mathcal{M}^{-1}\right) ^{ac}\varepsilon ^{cb}A_{\mu }\;.  \label{h}
\end{equation}%
The parameter $\gamma $ appearing in the previous expression has
the dimension of a mass and is called the Gribov parameter. It is
not a free parameter of the theory, being determined in a
self-consistent way through the gap equation
\cite{Capri:2005tj,Capri:2006cz,Capri:2008ak}
\begin{equation}
\frac{\delta \Gamma }{\delta \gamma ^{2}}=0\ .  \label{gbeq}
\end{equation}%
Therefore, for the partition function we write
\cite{Capri:2006cz,Capri:2008ak}
\begin{equation}
\mathcal{Z}=\int \mathcal{D}A\mathcal{D}b\mathcal{D}\bar{c}\mathcal{D}%
c\,e^{-\left( S_{\mathrm{YM}}+S_{\mathrm{MAG}}+S_{\mathrm{Hor}}\right) }\;.
\label{horizon}
\end{equation}%
\newline
\newline
The nonlocal term $S_{\mathrm{Hor}}$ can be localized by means of a pair of
complex vector bosonic fields, $(\phi _{\mu }^{ab},\bar{\phi}_{\mu }^{ab})$
according to
\begin{equation}
e^{-S_{\mathrm{Hor}}}=\int \mathcal{D}\bar{\phi}\mathcal{D}\phi \,\left(
\det \mathcal{M}\right) ^{8}\,\exp \left\{ -\int {d^{4}\!x\,}\left[ \,\bar{%
\phi}_{\mu }^{ab}\mathcal{M}^{ac}\phi _{\mu }^{ab}+\gamma ^{2}g\varepsilon
^{ab}\left( \phi _{\mu }^{ab}-\bar{\phi}_{\mu }^{ab}\right) A_{\mu }\,\right]
\right\} \;,  \label{eloc}
\end{equation}%
where the determinant $\left( \det \mathcal{M}\right) ^{8}$ takes into
account the Jacobian arising from the integration over the fields $(\phi
_{\mu }^{ab},\bar{\phi}_{\mu }^{ab})$. This term can also be localized by
means of a pair of complex vector anticommuting fields $(\omega _{\mu }^{ab},%
\bar{\omega}_{\mu }^{ab})$, namely

\begin{equation}
\left( \det \mathcal{M}\right) ^{8}=\int \mathcal{D}\bar{\omega}\mathcal{D}%
\omega \,\exp \left( \,\int {d^{4}\!x\,}\bar{\omega}_{\mu }^{ab}\mathcal{M}%
^{ac}\omega _{\mu }^{cb}\,\right) \;.  \label{det}
\end{equation}%
Moreover, as done in the case of the Landau gauge
\cite{Zwanziger:1989mf,Zwanziger:1992qr}, it will be useful to
perform the following shift in the variable $\omega _{\mu }^{ab}$
\cite{Capri:2006cz,Capri:2008ak}
\begin{equation}
\omega _{\mu }^{ab}\rightarrow \omega _{\mu }^{ab}+\left( \mathcal{M}%
^{-1}\right) ^{ac}\,\left( \mathcal{F}^{cd}\phi _{\mu }^{db}\right) \;,
\label{shift}
\end{equation}%
where the expression $\mathcal{F}^{ab}$ stands for
\begin{equation}
\mathcal{F}^{ab}=2g\varepsilon ^{ac}(\partial _{\mu }c+g\varepsilon
^{de}A_{\mu }^{d}c^{e})D_{\mu }^{cb}+g\varepsilon ^{ab}\partial _{\mu
}(\partial _{\mu }c+g\varepsilon ^{cd}A_{\mu }^{c}c^{d})-g^{2}(\varepsilon
^{ac}\varepsilon ^{bd}+\varepsilon ^{ad}\varepsilon ^{bc})A_{\mu
}^{d}(D_{\mu }^{ce}c^{e}+g\varepsilon ^{ce}A_{\mu }^{e}c)\;.  \label{shexp}
\end{equation}%
Therefore, the nonlocal horizon function gives place to a local term $S_{%
\mathrm{Local}}$
\begin{eqnarray}
e^{-S_{\mathrm{Hor}}} &=&\int \mathcal{D}\bar{\phi}\mathcal{D}\phi \mathcal{D%
}\bar{\omega}]\mathcal{D}\omega \,e^{-S_{\mathrm{Local}}}\;,  \notag \\
S_{\mathrm{Local}} &=&\int {d^{4}\!x}\left[ {\,}\bar{\phi}_{\mu }^{ab}%
\mathcal{M}^{ac}\phi _{\mu }^{cb}-\bar{\omega}_{\mu }^{ab}\mathcal{M}%
^{ac}\omega _{\mu }^{cb}++\bar{\omega}_{\mu }^{ab}\mathcal{F}^{ac}\phi _{\mu
}^{cb}+\gamma ^{2}g\varepsilon ^{ab}\left( \phi _{\mu }^{ab}-\bar{\phi}_{\mu
}^{ab}\right) A_{\mu }\right] \,\;,  \label{lact}
\end{eqnarray}%
so that we end up with a completely local action $S$ implementing the
restriction to the Gribov region $\Omega $, namely

\begin{equation}
S=S_{\mathrm{YM}}+S_{\mathrm{MAG}}+S_{\mathrm{Local}}\ .  \label{sloc}
\end{equation}%
Let us investigate now if the action $S$ displays exact $BRST$
invariance.  Following the analysis done in \cite{Dudal:2008sp},
let us first consider the case in which the Gribov parameter
$\gamma $ is set to zero, $\gamma =0$. In
this case, the action $S$ reduces to $S_{0}$%
\begin{eqnarray}
S_{0} &=&S_{\mathrm{YM}}+S_{\mathrm{MAG}}+\left. S_{\mathrm{Local}%
}\right\vert _{\gamma =0}\ ,  \notag \\
\left. S_{\mathrm{Local}}\right\vert _{\gamma =0} &=&\ \int {d^{4}\!x}\left[
{\,}\bar{\phi}_{\mu }^{ab}\mathcal{M}^{ac}\phi _{\mu }^{cb}-\bar{\omega}%
_{\mu }^{ab}\mathcal{M}^{ac}\omega _{\mu }^{cb}++\bar{\omega}_{\mu }^{ab}%
\mathcal{F}^{ac}\phi _{\mu }^{cb}\right] \ ,  \label{snot}
\end{eqnarray}%
which corresponds to the case in which the restriction to the
Gribov region has not been implemented. The physical content of
the action $S_{0}$ is thus the same as that of the Faddeev-Popov
action $S_{\text{\textrm{FP}}}$, eq.( \ref{zero}). In fact, it is
easily established that integration over the auxiliary fields
$(\phi _{\mu }^{ab},\bar{\phi}_{\mu }^{ab},\omega _{\mu
}^{ab},\bar{\omega}_{\mu }^{ab})$ amounts to introduce a unity
factor in the partition function. We expect thus that in this
case, the action $S_{0}$ displays exact $BRST\ $invariance. In
fact, introducing the following nilpotent $BRST\ $transformations
of the auxiliary fields $(\phi _{\mu }^{ab},\bar{\phi}_{\mu
}^{ab},\omega _{\mu }^{ab},\bar{\omega}_{\mu }^{ab})$

\begin{equation}
\begin{tabular}{cclccl}
$s\phi _{\mu }^{ab}$ & $\!\!\!=\!\!\!$ & $\omega _{\mu }^{ab}\;,\qquad $ & $%
s\omega _{\mu }^{ab}$ & $\!\!\!=\!\!\!$ & $0\;,\vspace{5pt}$ \\
$s\bar{\omega}_{\mu }^{ab}$ & $\!\!\!=\!\!\!$ & $\bar{\phi}_{\mu }^{ab}\;,$
& $s\bar{\phi}_{\mu }^{ab}$ & $\!\!\!=\!\!\!$ & $0\;,\vspace{5pt}$%
\end{tabular}
\label{brst_local}
\end{equation}%
it is easily checked that $\left. S_{\mathrm{Local}}\right\vert _{\gamma =0}$
can be cast in the form of an exact $BRST\ $variations
\begin{equation}
\left. S_{\mathrm{Local}}\right\vert _{\gamma =0}=\ s\int {d^{4}\!x\ }\left(
{\,}\bar{\omega}_{\mu }^{ab}\mathcal{M}^{ac}\phi _{\mu }^{cb}\right) \ ,
\label{sv}
\end{equation}%
so that $S_{0}$ displays exact $BRST$\ invariance,\textit{\ i.e.}
\begin{equation}
sS_{0}=0\ .  \label{einv}
\end{equation}%
It is worth remarking here that the auxiliary fields $(\phi _{\mu }^{ab},%
\bar{\phi}_{\mu }^{ab},\omega _{\mu }^{ab},\bar{\omega}_{\mu
}^{ab})$ transform in such a way that the nilpotency of the $BRST$
operator is preserved. In particular, from eq.(\ref{brst_local}),
it follows that these fields are assembled in $BRST$ doublets
\cite{Piguet:1995er}. As such, they do not alter the cohomology of
the operator $s$, which is identified by the colorless gauge
invariant operators built up with the field strength and its
covariant derivatives. In other words, the introduction of the
auxiliary fields does not modify the set of observables of the the
theory. \\\\Let us now consider the case in which $\gamma \neq 0$,
corresponding to the implementation of the restriction to the
Gribov region $\Omega $. As one can easily checks, the action $S$
of expression (\ref{sloc}) is not left
invariant by the $BRST$ transformations, eqs.(\ref{brst_fields}),(\ref%
{brst_local}). Instead one has the softly broken identity
\begin{equation}
sS=sS_{\mathrm{Local}}=\gamma ^{2}\Delta _{\gamma }\ ,  \label{sbr}
\end{equation}%
where $\Delta _{\gamma }$ is a dimension two soft breaking term, given by
\begin{equation}
\Delta _{\gamma }=g\int d^{4}x\;\left( \varepsilon ^{ab}\omega
_{\mu }^{ab}A_{\mu }-\varepsilon ^{ab}\left( \phi _{\mu
}^{ab}-\bar{\phi}_{\mu }^{ab}\right) (\partial _{\mu
}c+g\varepsilon ^{mn}A_{\mu }^{m}c^{n})\right) \ .  \label{sbt}
\end{equation}%
We see thus that, as in the case of the Landau gauge
\cite{Dudal:2008sp}, the restriction to the Gribov region $\Omega
$ entails a soft breaking of the $BRST\ $invariance. Notice that
the right hand side of eq.(\ref{sbr}) is proportional to the
Gribov parameter $\gamma $. The presence of this breaking is,
however, not unexpected. Its origin relies on the properties of
the Gribov region, being a consequence of the fact that
infinitesimal gauge transformations of field configurations
belonging to $\Omega $ give rise to configurations which are
located outside of $\Omega $. Therefore, the existence of a soft
breaking of the $BRST$ invariance looks rather natural. As already
underlined in \cite{Dudal:2008sp,Baulieu:2008fy}, the presence of
this breaking ensures that the Gribov parameter $\gamma $ is a
physical relevant parameter of the theory, entering the expression
of the gauge invariant correlation functions. This follows by
noting that
\begin{equation}
s\frac{\partial S}{\partial \gamma ^{2}}=\Delta _{\gamma }\ ,  \label{m1}
\end{equation}%
so that the expression $\left( \partial S/\partial \gamma ^{2}\right) $
cannot be written in the form of an exact $BRST$ term, namely%
\begin{equation}
\frac{\partial S}{\partial \gamma ^{2}}\neq s\widehat{\Delta }_{\gamma }\ ,
\label{m2}
\end{equation}%
for some local $\widehat{\Delta }_{\gamma }$. Equation (\ref{m2}) expresses
precisely the fact that $\gamma $ is not a gauge parameter of the theory.
The $BRST$ soft breaking is necessary in order to ensure that $\gamma $ is a
physical parameter of the theory. Suppose in fact that, instead of giving
rise to a soft breaking, the term $S_{\mathrm{Local}}$ would be left
invariant by the $BRST$ operator, \textit{i.e.}
\begin{equation}
sS_{\mathrm{Local}}\text{ }=0\;.  \label{a13}
\end{equation}%
Therefore, owing to the doublet structure of the auxiliary fields $(\phi
_{\mu }^{ab},\bar{\phi}_{\mu }^{ab},\omega _{\mu }^{ab},\bar{\omega}_{\mu
}^{ab})$, a local functional $\widehat{S}$ should exist such that
\begin{equation}
S_{\mathrm{Local}}=s\widehat{S}\;,  \label{a14}
\end{equation}%
from which it would follow that
\begin{equation}
\frac{\partial S_{\mathrm{Local}}}{\partial \gamma ^{2}}=s\frac{\partial
\widehat{S}}{\partial \gamma ^{2}}\;,  \label{a15}
\end{equation}%
which would imply that $\gamma ^{2}$ is an unphysical parameter.
\\\\As in the case of the Landau gauge
\cite{Zwanziger:1989mf,Zwanziger:1992qr,Maggiore:1993wq,Dudal:2005na,Dudal:2008sp},
the presence of the soft breaking term, eq.(\ref{sbr}), does not
spoil the renormalizability of the theory
\cite{Capri:2006cz,Capri:2008ak}. This remarkable feature relies
on the possibility of extending to the maximal Abelian gauge the
same procedure outlined by Zwanziger in the case of the Landau
gauge \cite{Zwanziger:1989mf,Zwanziger:1992qr}, amounting to embed
$S_{\mathrm{Local}}$ into a generalized action,
$S_{\mathrm{Local}}^{\mathrm{inv}}$, which enjoys exact BRST
invariance, namely
\begin{equation}
S_{\mathrm{Local}}\rightarrow S_{\mathrm{Local}}^{\mathrm{inv}}\;,\qquad sS_{%
\mathrm{Local}}^{\mathrm{inv}}=0\;.  \label{m3}
\end{equation}%
Moreover, the original action action $S_{\mathrm{Local}}$ can be
recovered from the generalized action
$S_{\mathrm{Local}}^{\mathrm{inv}}$ by demanding that some
external sources of $S_{\mathrm{Local}}^{\mathrm{inv}}$ acquire a
particular value. Let us elaborate more on this point. Following
\cite{Capri:2006cz,Capri:2008ak}, the generalized $BRST$ invariant
action $S_{\mathrm{ Local}}^{\mathrm{inv}}$ turns out to be given
by the expression
\begin{eqnarray}
S_{\mathrm{Local}}^{\mathrm{inv}} &=&s\int {d^{4}\!x\,}\Bigl(\bar{\omega}%
_{\mu }^{ab}\mathcal{M}^{ac}\phi _{\mu }^{cb}-\bar{N}_{\mu \nu }^{ab}D_{\mu
}^{ac}\phi _{\nu }^{cb}+M_{\mu \nu }^{ab}D_{\mu }^{ac}\bar{\omega}_{\nu
}^{cb}\Bigr)  \notag \\
&=&\int {d^{4}\!x\,}\Bigl\{\bar{\phi}_{\mu }^{ab}\mathcal{M}^{ac}\phi _{\mu
}^{cb}-\bar{\omega}_{\mu }^{ab}\mathcal{M}^{ac}\omega _{\mu }^{cb}+\bar{%
\omega}_{\mu }^{ab}\mathcal{F}^{ac}\phi _{\mu }^{cb}+\bar{M}_{\mu \nu
}^{ab}D_{\mu }^{ac}\phi _{\nu }^{cb}+N_{\mu \nu }^{ab}D_{\mu }^{ac}\bar{%
\omega}_{\nu }^{cb}  \notag \\
&&+\bar{N}_{\mu \nu }^{ab}[\,D_{\mu }^{ac}\omega _{\nu }^{cb}+g\varepsilon
^{ac}(\partial _{\mu }c+g\varepsilon ^{de}A_{\mu }^{d}c^{e})\phi _{\nu
}^{cb}\,]+M_{\mu \nu }^{ab}[\,D_{\mu }^{ac}\bar{\phi}_{\nu
}^{cb}+g\varepsilon ^{ac}(\partial _{\mu }c+g\varepsilon ^{de}A_{\mu
}^{d}c^{e})\bar{\omega}_{\nu }^{cb}\,]\Bigr\}\;,  \label{S_inv}
\end{eqnarray}%
where $(M_{\mu \nu }^{ab},\bar{M}_{\mu \nu }^{ab})$, $(N_{\mu \nu }^{ab},%
\bar{N}_{\mu \nu }^{ab})$ are external sources transforming as $BRST\ $%
doublets, \textit{i.e}.
\begin{equation}
\begin{tabular}{cclccl}
$sM_{\mu \nu }^{ab}$ & $\!\!\!=\!\!\!$ & $N_{\mu \nu }^{ab}\,,\qquad $ & $%
sN_{\mu \nu }^{ab}$ & $\!\!\!=\!\!\!$ & $0\,,\vspace{5pt}$ \\
$s\bar{N}_{\mu \nu }^{ab}$ & $\!\!\!=\!\!\!$ & $-\bar{M}_{\mu \nu }^{ab}\,,$
& $s\bar{M}_{\mu \nu }^{ab}$ & $\!\!\!=\!\!\!$ & $0\,.$%
\end{tabular}
\label{m4}
\end{equation}%
In order to recover $S_{\mathrm{Local}}$ from the $BRST$ invariant action $%
S_{\mathrm{Local}}^{\mathrm{inv}}$ we first take the physical
limit of the external sources $(M_{\mu \nu }^{ab},\bar{M}_{\mu \nu
}^{ab})$, $(N_{\mu \nu }^{ab},\bar{N}_{\mu \nu }^{ab})$, which is
defined by \cite{Capri:2006cz,Capri:2008ak}
\begin{equation}
\begin{tabular}{cclcl}
$M_{\mu \nu }^{ab}\Bigl|_{\mathrm{phys}}$ & $\!\!\!=\!\!\!$ & $-\bar{M}_{\mu
\nu }^{ab}\Bigl|_{\mathrm{phys}}$ & $\!\!\!=\!\!\!$ & $-\delta ^{{ab}%
}{}\delta _{{\mu \nu }}\gamma ^{2}\;,\vspace{5pt}$ \\
$N_{\mu \nu }^{ab}\Bigl|_{\mathrm{phys}}$ & $\!\!\!=\!\!\!$ &
$-\bar{N}_{\mu \nu }^{ab}\Bigl|_{\mathrm{phys}}$ & $\!\!\!=\!\!\!$
& $0\;,$ \label{physvalue}
\end{tabular}
\end{equation}
and then we perform a shift in the variable $\omega _{\mu }^{ab}$
as \cite{Capri:2006cz,Capri:2008ak}
\begin{equation}
\omega _{\mu }^{ab}\rightarrow \omega _{\mu }^{ab}+\left( \mathcal{M}%
^{-1}\right) ^{ac}\left[ \,\gamma ^{2}g\varepsilon ^{cb}(\partial _{\mu
}c+g\varepsilon ^{de}A_{\mu }^{d}c^{e})\right] \;,  \label{m6}
\end{equation}%
so that
\begin{equation}
S_{\mathrm{Local}}^{\mathrm{inv}}\Bigl|_{\mathrm{phys}}=S_{\mathrm{Local}}\;.
\label{m7}
\end{equation}%
Let us conclude by mentioning that the possibility of writing down
the generalized action $S_{\mathrm{Local}}^{\mathrm{inv}}$ enables
us to obtain generalized Slavnov-Taylor identities
\cite{Capri:2006cz,Capri:2008ak} which can be used to establish
the renormalizability of the generalized action $S_{\mathrm{Local}}^{\mathrm{inv}%
}$ and, in particular, of the action $S$, eq.(\ref{sloc}).

\subsection{The U(1) local Ward identity}

As we have seen in the previous section, the soft breaking of the BRST
invariance is deeply related to the properties of the Gribov region and, in
particular, to the existence of a boundary $\partial \Omega $ along the
off-diagonal directions. One should also notice that, from eqs.(\ref{gauge}%
), it follows that, when restricted to the diagonal direction,
amounting to set to zero the off-diagonal parameters $\omega
^{a}$, the gauge transformations take the form
\begin{eqnarray}
\delta _{\mathrm{diag}}A_{\mu } &=&-\partial _{\mu }{\omega }\ ,  \notag \\
\;\delta _{\mathrm{diag}}A_{\mu }^{a} &=&-g\varepsilon ^{ab}A_{\mu
}^{b}\omega \;,  \label{p1}
\end{eqnarray}%
where ${\omega }$ is the diagonal parameter corresponding to the $U(1)$
Cartan subgroup. From eqs.(\ref{p1}) one sees that the diagonal field $%
A_{\mu }$ transforms as an Abelian $U(1)$ gauge field, while the
off-diagonal components $A_{\mu }^{a}$ play the role of charged matter
fields. Moreover, since the Gribov region is unbounded along the diagonal
direction, transformations (\ref{p1}) are expected to correspond to an
invariance of the action $S$ of eq.(\ref{sloc}). In other words, expression (%
\ref{sloc}) should display a local $U(1)$ Ward identity, and this
in the presence of the horizon term $S_{\mathrm{Hor}}$,
eq.(\ref{h}). This turns out to be the case. In fact, the action
$S$ enjoys the $U(1)$ local Ward identity
\begin{equation}
\partial _{\mu }\frac{\delta S}{\delta A_{\mu }}+
g\varepsilon ^{ab}\sum_{{\Phi }}\Phi ^{a}\frac{\delta S}{\delta \Phi ^{b}}=-i\partial
^{2}b\;,  \label{pw}
\end{equation}%
where we have set $\Phi =\left( A_{\mu }^{a},b^{a},\bar{c}^{a},c^{a},\bar{%
\phi}_{\mu }^{ab},\phi _{\mu }^{ab},\bar{\omega}_{\mu }^{ab},\omega _{\mu
}^{ab}\right) $ for all off-diagonal fields. The existence of the local $%
U(1) $ Ward identity is an important feature of the maximal
Abelian gauge, supporting the so called Abelian dominance. It
nicely fits with the unboundedness of the Gribov region $\Omega $
in the diagonal direction.

\section{An overview on the gluon and ghost propagators}

This section provides a short summary of the main results which
have been obtained on the gluon and ghost propagators when taking
into account the restriction to the Gribov region
\cite{Capri:2006cz,Capri:2008ak}. Let us spend first a few words
on dimension two condensates. One should notice that the
introduction of the horizon function $S_{\mathrm{Hor}}$ in its
localized form, expression (\ref{lact}), entails the introduction
of a dimension two condensate. In fact, the gap equation
(\ref{gbeq}), implies that the dimension two operator $\left(
\varepsilon ^{ab}\left( \phi _{\mu }^{ab}-\bar{\phi}_{\mu
}^{ab}\right) A_{\mu }\right) $ acquires a nonvanishing
expectation value, \textit{i.e.} $\left\langle \varepsilon
^{ab}\left( \phi _{\mu }^{ab}-\bar{\phi}_{\mu }^{ab}\right) A_{\mu
}\right\rangle $ $\neq 0$. An analogous condensate is found in the
Landau gauge
\cite{Zwanziger:1989mf,Zwanziger:1992qr,Dudal:2005na,Dudal:2007cw,Dudal:2008sp},
where the gap equation for the Gribov parameter $\gamma $ implies
that $\left\langle f^{ABC}\left( \phi _{\mu }^{AB}-\bar{\phi}_{\mu
}^{AB}\right) A_{\mu }^{C}\right\rangle \neq 0$, where $\left(
\phi _{\mu }^{AB},\bar{\phi}_{\mu }^{AB}\right) $ are the
auxiliary fields needed for the localization of the horizon
function in the Landau gauge and the indices $A,B,C$ belong to the
adjoint representation of $SU(N)$, $A,B,C=1,...,N^{2}-1$. \\\\
Furthermore, in complete analogy with the case of the Landau gauge
\cite{Dudal:2005na,Dudal:2007cw,Dudal:2008sp}, other dimension two
condensates have to be taken into account in the maximal Abelian
gauge. More precisely, the following dimension two operators can
be introduced in a way which preserves renormalizability of the
theory as well as its symmetry content \cite{Capri:2008ak}:
\begin{equation}
\mathcal{O}_{A^{2}}=\,A_{\mu }^{a}A_{\mu }^{a}\ ,  \label{c1}
\end{equation}%
\begin{equation}
\mathcal{O}_{\mathrm{ghost}}=g\varepsilon ^{ab}\bar{c}^{a}c^{b},  \label{c2}
\end{equation}
\begin{equation}
\mathcal{O}_{\bar{f}f}=(\bar{\phi}_{\mu }^{ab}\phi _{\mu }^{ab}-\bar{\omega}%
_{\mu }^{ab}\omega _{\mu }^{ab}-\bar{c}^{a}c^{a})\;.  \label{c3}
\end{equation}%
The operator (\ref{c1}) is related to the dynamical mass
generation for off-diagonal gluons, a feature which supports the
Abelian dominance hypothesis. Its condensation has been
established in \cite{Dudal:2004rx}, where a dynamical off-diagonal
gluon mass $m\simeq 2.2\Lambda _{\overline{MS}}$ \ has been
reported. The ghost operator (\ref{c2}) is needed in order to
account for the dynamical breaking of the $SL(2,R)$ symmetry
present in the ghost sector of the maximal Abelian gauge. Its
condensation has been analysed recently in \cite{Capri:2007hw}.
Concerning the third operator, eq.(\ref{c3}), we notice that it
depends on the auxiliary fields $\left( \bar{\phi}_{\mu
}^{ab},\phi _{\mu }^{ab},\bar{\omega}_{\mu }^{ab},\omega _{\mu
}^{ab}\right) $. It is in fact needed to account for the
nontrivial dynamics developed by those fields \cite{Capri:2008ak}.
An analogous operator has been found in the Landau gauge
\cite{Dudal:2005na,Dudal:2007cw,Dudal:2008sp}, where it allows to
reconcile the Gribov-Zwanziger framework with the most recent
lattice data on the gluon and ghost propagators
\cite{Cucchieri:2008fc,Cucchieri:2007rg}. Let us now summarize our
results \cite{Capri:2008ak} on the tree level gluon and ghost
propagators:

\begin{itemize}
\item {\textbf{The off-diagonal gluon propagator:} \newline
the transverse off-diagonal gluon propagator turns out to be of the Yukawa
type
\begin{equation}
\langle A_{\mu }^{a}(-k)A_{\nu }^{b}(k)\rangle =\frac{1}{k^{2}+m^{2}}\left(
\delta _{\mu \nu }-\frac{k_{\mu }k_{\nu }}{k^{2}}\right) \delta ^{ab}\;,
\label{dg}
\end{equation}%
where $m$ is the dynamical mass originating from the condensation of the
gluon operator (\ref{c1}) } \noindent \newline

\item {\textbf{The diagonal gluon propagator:} \newline
for the diagonal gluon propagator we have obtained an infrared suppressed
propagator of the Gribov-Stingl type, namely
\begin{equation}
\langle A_{\mu }(-k)A_{\nu }(k)\rangle =\frac{k^{2}+\mu ^{2}}{k^{4}+\mu
^{2}k^{2}+4\gamma ^{4}g^{2}}\left( \delta _{\mu \nu }-\frac{k_{\mu }k_{\nu }%
}{k^{2}}\right) \;,  \label{diaggluon}
\end{equation}%
where $\gamma $ is the Gribov parameter and $\mu $ is a mass parameter
related to the condensation of the operator (\ref{c3}). } We observe that
expression {(\ref{diaggluon})} does not vanish at the origin. It gives rise
to a positivity violating propagator in configuration space, a feature
usually interpreted as evidence for gluon confinement. \newline

\item {\textbf{The symmetric off-diagonal ghost propagator:}}
\newline for the symmetric off-diagonal ghost propagator we have
found
\begin{equation}
\langle \bar{c}^{a}(-k)c^{b}(k)\rangle _{\mathrm{symm}}=\frac{k^{2}+\mu ^{2}%
}{k^{4}+2\mu ^{2}k^{2}+(\mu ^{4}+v^{4})}\,{\delta^{{ab}}}\;,
\label{symmghost}
\end{equation}%
where $v$ is a mass parameter related to the condensation of the
ghost operator (\ref{c2}). Notice that expression
{(\ref{symmghost})} is suppressed in the infrared and attains a
nonvanishing finite value at $k=0$. \newline

\item {\textbf{The antisymmetric off-diagonal ghost propagator:}}
\newline finally, for the antisymmetric off-diagonal ghost
propagator we have
\begin{equation}
\langle \bar{c}^{a}(-k)c^{b}(k)\rangle _{\mathrm{antisymm}}=\frac{v^{2}}{%
k^{4}+2\mu ^{2}k^{2}+(\mu ^{4}+v^{4})}\,\varepsilon ^{ab}\;.
\label{antisymmghost}
\end{equation}
As expected, this behavior is a consequence of the ghost
condensate \cite{Capri:2007hw}, $\langle \varepsilon
^{ab}\bar{c}^{a}c^{b}\rangle \sim v^{2}$.
\end{itemize}
It is worth mentioning that the behavior shown above for the gluon
and ghost propagators turns out to be in remarkable agreement with
the most recent lattice data, as reported in
\cite{Mendes:2006kc,Mendes:2008ux}.

\section{Conclusion}

In this work a study of the Gribov region $\Omega $ in the maximal
Abelian gauge has been performed. Several features of this region
have been established. The region $\Omega $ has been proven to be
bounded in all off-diagonal directions, while it turns out to be
unbounded in the diagonal one. The convexity of $\Omega $ has also
been established. Roughly speaking, the region $\Omega \ $ looks
like an infinite cylinder along the diagonal axis in field space.
\\\\The Gribov region of the maximal Abelian gauge looks deeply
different from the corresponding region of the Landau gauge, which
is in fact bounded in all directions in field space. \\\\The
results which have been obtained give us a better understanding of
several features of the maximal Abelian gauge. This is the case of
the soft breaking of the BRST\ invariance, deeply related to the
fact that the region $\Omega $ turns out to be bounded in the
off-diagonal directions. Moreover, the unboundedness of $\Omega$
along the diagonal direction is at the origin of the local $U(1)$
Ward identity (\ref{pw}). \\\\We also point out that our results
nicely fit within the hypothesis of the Abelian dominance, which
is a key ingredient for the dual superconductivity picture for
color confinement in the maximal Abelian gauge. The picture which
emerges from our analysis is that the relevant configurations in
the low energy nonperturbative region should be those located very
close the diagonal axis in field space. This is supported by the
following considerations:

\begin{itemize}
\item the ghost propagator,
eqs.{(\ref{symmghost}),(\ref{antisymmghost}), is non-singular and
attains a finite value at the origin }$k\simeq 0$, a result in
very good agreement with the lattice data
\cite{Mendes:2006kc,Mendes:2008ux}.  This suggests that the
relevant configurations  are not located near the boundary
$\partial \Omega $ of the Gribov region, where the ghost
propagator becomes divergent.

\item all lattice data obtained so far on the gluon propagators
\cite{Amemiya:1998jz,Bornyakov:2003ee,Mendes:2006kc,Mendes:2008ux}
give a clear indication of the fact that the off-diagonal gluon
propagator is of the Yukawa type, in agreement with expression
(\ref{dg}),  and turns out to be suppressed in the infrared with
respect to the diagonal gluon propagator which, moreover, can be
nicely fitted by a Gribov-Stingl propagator
\cite{Mendes:2006kc,Mendes:2008ux}, in remarkable agreement with
expression {(\ref{diaggluon}). }

\item finally, it has to be noted that when an Abelian
configuration in the maximal Abelian gauge, {\it i.e.} a
configuration lying on the diagonal axis in field space, is gauge
transformed so as to fulfill the Landau gauge condition, it is
mapped into a configuration lying on the horizon $\partial \Omega
$ of the Gribov region $\Omega $ of the Landau gauge
\cite{Greensite:2004ke,zw}. These configurations are believed to
play a relevant role for gluon confinement in the Landau gauge.
This observation might have profound consequences in order to
consistently relate our current understanding of the confinement
mechanism in different gauges.
\end{itemize}

\noindent Much work is still needed in order to unravel the
intricacies of the maximal Abelian gauge.\ Needless to say, a
characterization of the properties of the fundamental modular
region of the maximal Abelian gauge would be a very relevant
achievement, a task which is beyond our present capabilities.
Nevertheless, we hope that our present work will stimulate further
investigations on the maximal Abelian gauge.

\section*{Acknowledgments}

S. P. Sorella thanks D. Zwanziger for valuable discussions. The
Conselho Nacional de Desenvolvimento Cient\'{i}fico e
Tecnol\'{o}gico (CNPq-Brazil), the Faperj, Funda{\c{c}}{\~{a}}o de
Amparo {\`{a}} Pesquisa do Estado do Rio de Janeiro, the SR2-UERJ
and the Coordena{\c{c}}{\~{a}}o de Aperfei{\c{c}}oamento de
Pessoal de N{\'{i}}vel Superior (CAPES), the CLAF, Centro
Latino-Americano de F{\'\i}sica, are gratefully acknowledged for
financial support.

\end{document}